\documentclass[12pt,preprint]{aastex}

\shorttitle{Exoplanet Host Star Companions}
\shortauthors{Baines et al.}

\begin{document}

\title{The Search for Stellar Companions to Exoplanet Host Stars \\ Using the CHARA Array}

\author{Ellyn K. Baines, Harold A. McAlister, Theo A. ten Brummelaar, Nils~H.~Turner, Judit Sturmann \& Laszlo Sturmann}
\affil{Center for High Angular Resolution Astronomy, Georgia State University, P.O. Box 3969, Atlanta, GA 30302-3969}
\email{baines, hal@chara.gsu.edu; theo, nils, judit, sturmann@chara-array.org}

\author{Stephen T. Ridgway}
\affil{Kitt Peak National Observatory, National Optical Astronomy Observatory, P.O. Box 26732, Tucson, AZ 85726-6732} 
\email{ridgway@noao.edu}

\altaffiltext{}{For preprints, please email baines@chara.gsu.edu.}

\begin{abstract}
Most exoplanets have been discovered via radial velocity studies, which are inherently insensitive to orbital inclination. Interferometric observations will show evidence of a stellar companion if it sufficiently bright, regardless of the inclination. Using the CHARA Array, we observed 22 exoplanet host stars to search for stellar companions in low-inclination orbits that may be masquerading as planetary systems. While no definitive stellar companions were discovered, it was possible to rule out certain secondary spectral types for each exoplanet system observed by studying the errors in the diameter fit to calibrated visibilities and by searching for separated fringe packets.
\end{abstract}


\section{Introduction}
In studies of exoplanet systems, certain assumptions are made about the inclination ($i$) of the systems discovered; i.e., it is assumed the orbit has an intermediate to high inclination ($i \sim 45-90^\circ$) because the probability of the orbit being nearly face-on ($i \sim 0^\circ$) is extremely low. If we assume a random sample of orbital orientations along our line of sight, the probability of the inclination being less or equal to a given $i$ is [$1 - \cos (i)$] while the probability of an inclination being greater than a given $i$ is $\cos (i)$. Therefore the probability of an orbit with an inclination below 45$^\circ$ is $\sim$30\% while the probability of the orbit having an inclination above 45$^\circ$ is $\sim$70\%.

This conclusion implies that the calculated companion mass, known only as the quantity $m_{\rm p} \sin^3 i$ where $m_{\rm p}$ is the mass of the planet, is planetary in nature instead of stellar. While this is probably a safe conclusion for the majority of the exoplanets discovered, the chance remains that in a large enough sample, a few of the candidate planetary systems may be face-on binary star systems instead.

\begin{figure}[!t]
  \centering \includegraphics[angle=90,width=1.0\textwidth]
  {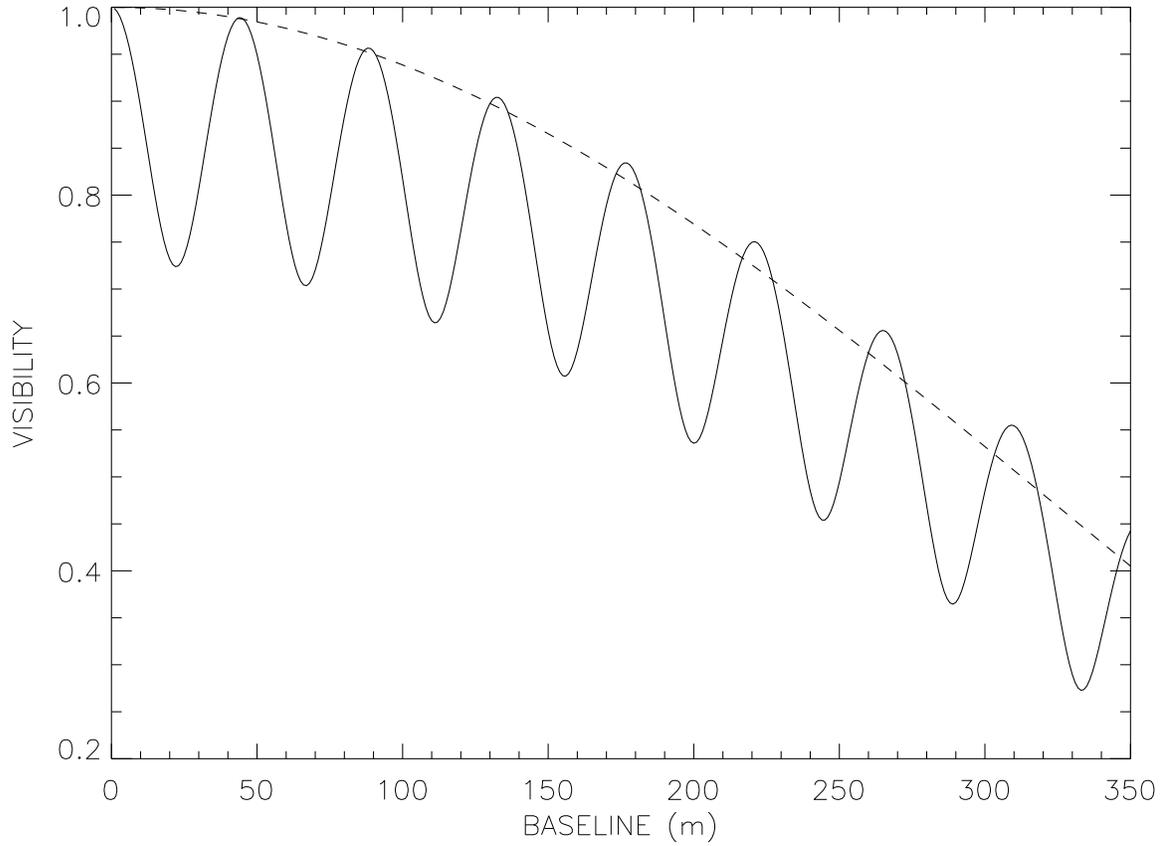}\\
 \caption{Example of the difference between the visibility curves for a single star and a binary system. The dotted line indicates the curve for a single star with $\theta$=1.0~mas, while the solid line represents the curve for a binary system with the following parameters: $\theta_{\rm primary}$=1.0~mas, $\theta_{\rm secondary}$=0.5~mas, $\alpha$=10~mas, and $\Delta K$=2.0.}
  \label{singlevsbinary_viscurve}
\end{figure}

We make no assumptions about the inclination. If a second star is present and is not more than $\sim$2.5 magnitudes fainter than the host star, the effects of the second star will be seen in the interferometric visibility curve\footnote{A limiting $\Delta K$ of 2.5 is a lower limit, as the true $\Delta K$ also depends on the absolute brightness of the two stars and could be higher for other systems.}. See Figure~\ref{singlevsbinary_viscurve} for an example of the difference between the visibility curves for a single star and binary system.

Two studies have shown that radial velocity observations of exoplanet systems alone are insufficient to distinguish between intermediate- to high-inclination planetary systems and low-inclination binary star systems. \citet{2001A&A...371..250S} estimated probability densities of orbital periods and eccentricities for a sample of exoplanet candidates and a sample of spectroscopic binary star systems with solar-type primary stars in order to determine if there were any fundamental differences between the two types of systems. They found the respective distributions of the two populations were statistically indistinguishable from each other in the context of orbital elements.

In an earlier study, \citet{1998A&A...334L..37I} modeled nine known exoplanet systems as binary star systems to test if the radial velocity observations could be explained by low-mass stellar companions. Although the probability of binary star systems appearing as planetary systems was low, ranging from 0.01 to 4\%, the model results described the observations satisfactorily and showed it is possible for a binary star system to mimic an exoplanet system.

\citet{2007ApJ...670..820W} argue that $\sim 2.5$\% of the exoplanet systems with Jupiter-mass planets with semimajor axes of $<$0.1 AU may have formed in binary systems where the binary star companion is highly inclined to the planet's orbit. Furthermore, it has been proposed that planets form in the same manner in both single-star and binary-star systems during certain stages \citep{2007ApJ...669.1316T}. This would be significant due to the fact that more than half of all stars form in binary or multiple star systems.

\section{Interferometric Observations}
Our target list was derived from the general exoplanet list using declination limits and magnitude constraints. The stars needed to be north of -10$^\circ$, brighter than $V=+10$ in order for the tip/tilt subsystem to lock onto the star, and brighter than $K=+6.5$ so fringes were easily visible. This reduced the original list to $\sim$80 targets, and we observed 22 systems from 2004 January to 2007 September.

The stars were observed using the Center for High Angular Resolution Astronomy (CHARA) Array, a six-element Y-shaped interferometric array located on Mount Wilson, California \citep{2005ApJ...628..453T}. The Array presently uses visible wavelengths at 470-800 nm for tracking and tip/tilt corrections and near infrared bands, $H$ at 1.67~$\mu$m and $K^\prime$ at 2.15~$\mu$m, for fringe detection and data collection. All observations of the host stars were obtained using the pupil-plane ``CHARA Classic'' beam combiner in the $K^\prime$-band.

The observations were taken using many of the baselines the CHARA Array offers from intermediate-length baselines at 108 m to the longest baseline available at 331 m. Table \ref{observations} lists the exoplanet host stars observed, their calibrators, the baselines used, the dates of the observations, and the number of observations obtained. Calibrators were chosen to have small predicted angular diameters in order to reduce errors in the target's angular diameter measurement, and the calibrators were as close to the target stars as possible, usually within 5$^\circ$. This allowed for less time between calibrator and target observations, thus reducing the effects of changing seeing conditions as much as possible. 

We observed using the standard calibrator-target-calibrator pattern, which allowed us to see if the target star was changing with respect to the calibrator over time, assuming the calibrator did not have an unseen stellar companion or a circumstellar disk. In order to select reliable calibrators, we searched the literature for indications of binarity and performed spectral energy distribution fits to published $UBVRIJHK$ photometry to see if there was any excess flux that would indicate a stellar companion. Calibrator candidates with variable radial velocities were discarded even if their SEDs displayed no characteristics of duplicity.

Stellar companions in low-inclination orbits could show themselves in two ways: in the residuals to the diameter fit to calibrated visibilities (sensitive to companions at separations of $\sim$0.5 to 10 milliarcseconds) and as separated fringe packets (for companions in the separation range of $\sim$10 to 50 milliarcseconds). The two methods are described in more detail below.


\section{Characterizing Residuals to Diameter Fit}
\begin{figure}[!t]
  \centering \includegraphics[angle=90,width=1.0\textwidth]
  {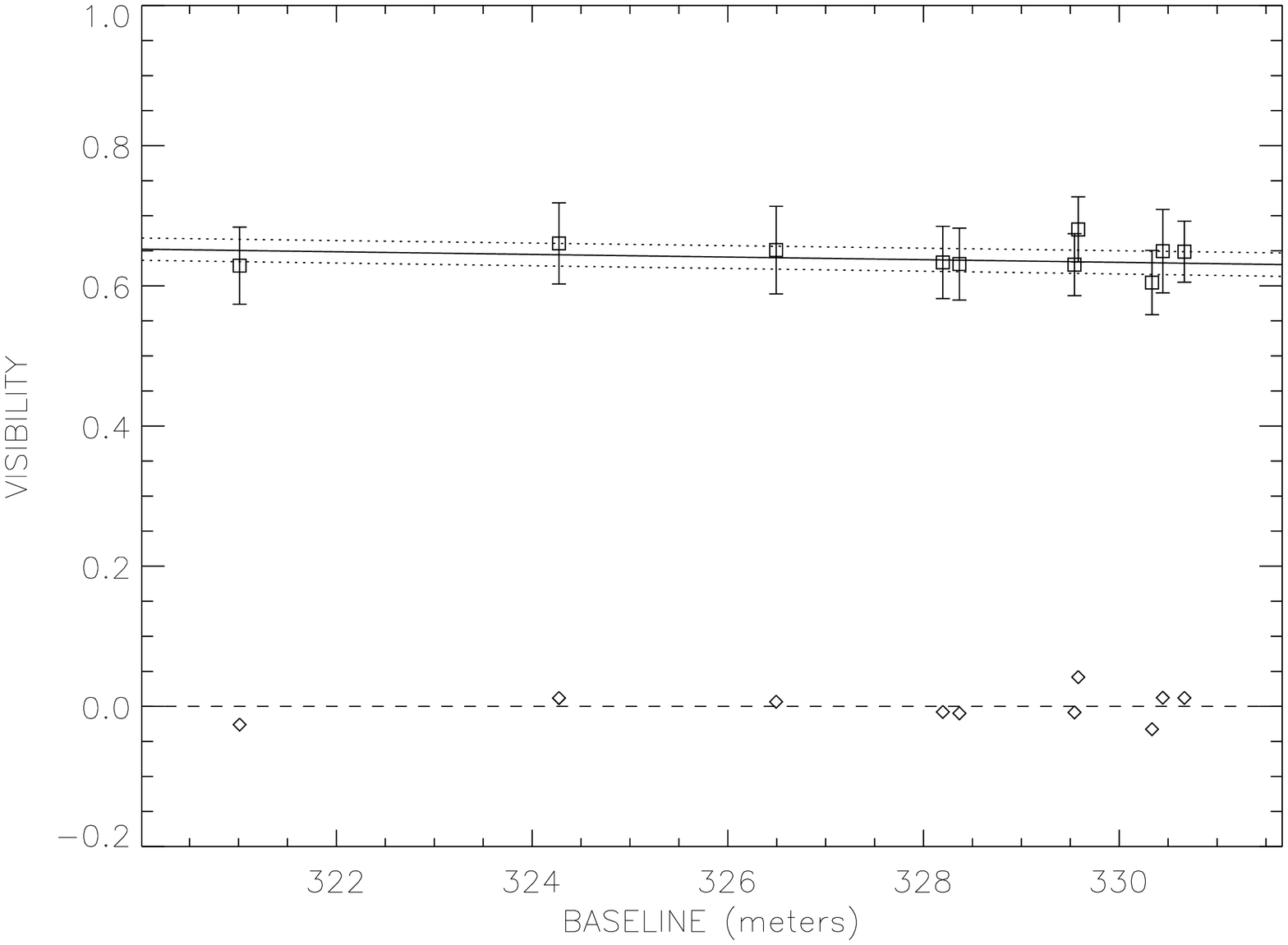}\\
 \caption{Example of residuals indicative of a single star. The solid line represents the theoretical visibility curve for a star with the best fit diameter, the dashed lines are the 1$\sigma$ error limits of the diameter fit, the $\Box$s are the calibrated visibilities ($V_c$), the vertical lines are the errors in $V_c$, and the $\Diamond$s are the residuals to the diameter fit. Note how the residuals fall evenly both above and below the line at $V = 0$ and show no sinusoidal trends. (Data for HD~120136.)}
  \label{exp_goodresid}
\end{figure}

\begin{figure}[!t]
  \centering \includegraphics[angle=90,width=1.0\textwidth]
  {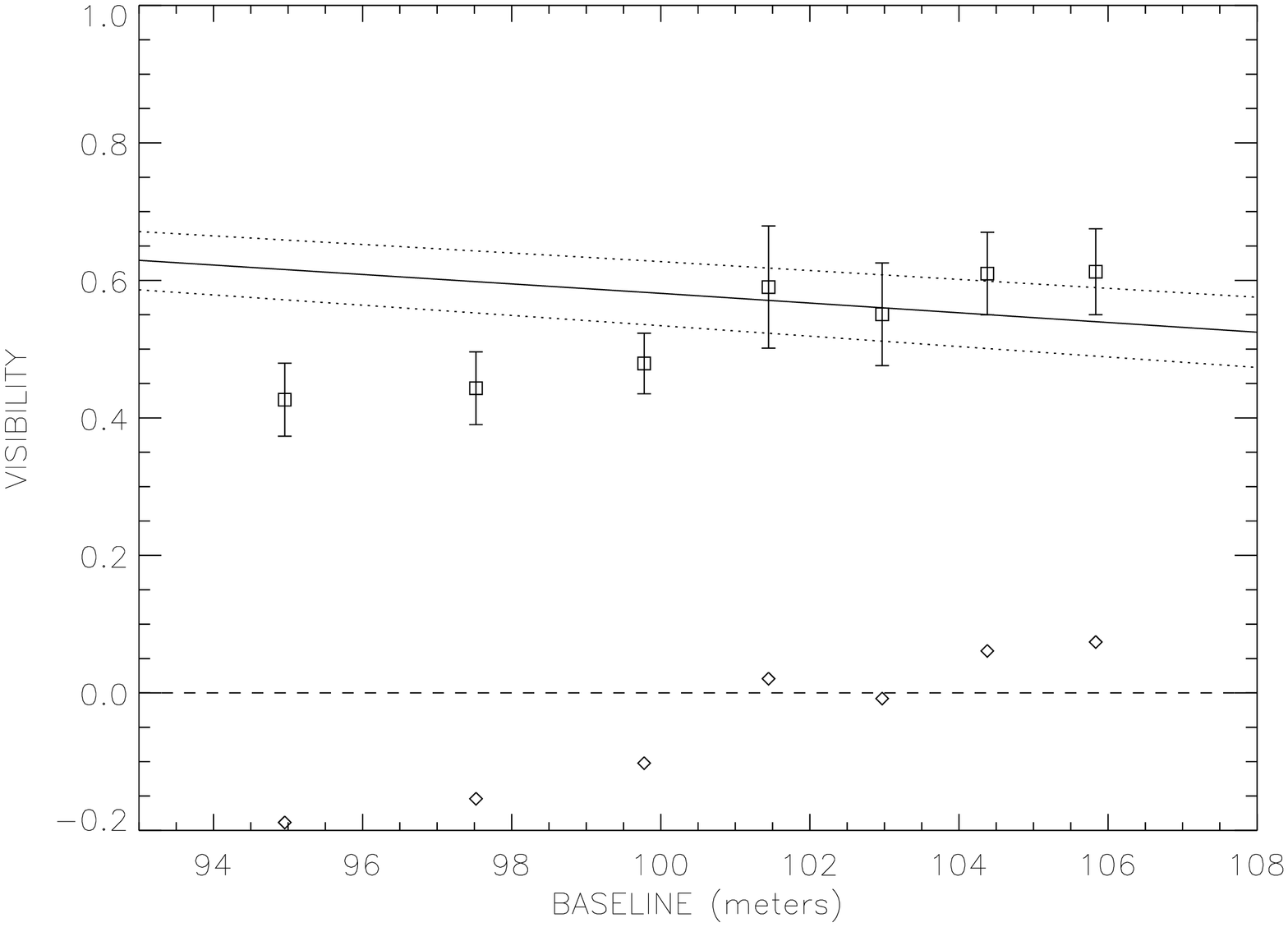}\\
  \caption{Example of residual systematics indicative of a binary star. The solid line represents the theoretical visibility curve for a star with the best fit diameter, the dashed lines are the 1$\sigma$ error limits of the diameter fit, the $\Box$s are the calibrated visibilities ($V_c$), the vertical lines are the errors in $V_c$, and the $\Diamond$s are the residuals to the diameter fit. Note how the residuals are systematically low for the shorter-baseline observations and systematically high for the longer-baseline observations. (Data for $\beta$~Aurigae.)}
  \label{exp_oddresid}
\end{figure}

The systematics in the residuals of the diameter fit to measured visibilities indicate whether or not a stellar companion may be present. For a single star, the residuals show a Gaussian distribution about 0 (see Figure \ref{exp_goodresid}). On the other hand, Figure \ref{exp_oddresid} shows the strongly systematic behavior in the residuals for a known binary star system. Inspection of Figure \ref{exp_goodresid} shows that the error estimates of the calibrated visibilities in all cases exceed the (observed - calculated) residuals ($\sigma_{\rm res}$). As described in \citet{2005ApJ...628..439M}, the error estimates in the measured instrumental visibilities are conservatively taken from the standard deviation from the mean of subgroupings of individual visibility scans. In this case, those estimates provide a reduced $\chi^2$ less than unity, which indicates that our individual visibility error estimates are too large.

The standard deviation for the residuals for Figure \ref{exp_goodresid} is $\pm$0.056 while $\sigma_{\rm res}$~=~$\pm$0.106 for Figure \ref{exp_oddresid}. The typical $\sigma_{\rm res}$ for a single star is well under $\pm$0.100 while $\sigma_{\rm res}$ for a binary with a low $\Delta K$ is usually $\gtrsim \pm$0.100. For example, the binary system HD~146361, which is composed of two nearly-identical stars \citep{2003A&A...399..315S} and has a $\Delta K \sim$ 0.2, was observed by Deepak Raghavan for three nights in May of 2007 along with the calibrator star HD~152598. The measured $\sigma_{\rm res}$ for the three nights observations are $\pm$0.180, $\pm$0.229, and $\pm$0.118 indicating a departure from the single-star model for this system (private communication). 

Another diagnostic to distinguish between the single star seen in Figure 2 and the binary star system shown in Figure 3 is to compare their $\chi^2$ values. A low $\chi^2$ value indicates a good fit to the single-star model while a large value indicates a poor fit, revealing the presence of a secondary star, circumstellar disk, or an asymmetry of the star. For example, the $\chi^2$ for the star shown in Figure 2 is 1.80 while $\chi^2$ = 20.89 for the binary star seen in Figure 3.

For each exoplanet system, a variety of low-mass secondary stars were considered: G5~V, K0~V, K5~V, M0~V, and M5~V. Most of the stars in the sample are solar-type stars, so more massive main sequence spectral types need not be considered. The magnitude difference ($\Delta$M$_K$, listed as $\Delta K$ in the tables) and angular separation ($\alpha$) of a face-on orbit between the host star and companion were calculated for each possible pairing:

\begin{equation}
\Delta \mathrm{M}_K = \mathrm{M}_{\rm s} + (\mathrm{m}_{\rm h} - \mathrm{M}_{\rm h}) - \mathrm{m}_{\rm h},
\end{equation}
where M$_{\rm h,s}$ are the absolute magnitudes of the host star and potential secondary, respectively, m$_{\rm h}$ is the apparent magnitude of the host star, and
\begin{equation}
\rm (m_{h} - M_{h}) = 5 \; log \left( \frac{100}{\pi} \right),
\end{equation}
where $\pi$ is the host star's parallax in milliarcseconds (mas). An estimate of the angular separation $\alpha$ in mas was calculated from Kepler's Third Law and $\pi$:
\begin{equation}
\alpha = \left[ \left(M_{\rm h}+M_{\rm s} \right) \times P^2 \right]^{\frac{1}{3}} \times \pi ,
\end{equation}
where $M_{\rm h,s}$ are the masses (in $M_\odot$) of the exoplanet's host star and potential secondary star, respectively, and $P$ is the companion's orbital period in years.

Additionally, the angular diameter $\theta$ of the possible secondary was estimated using the calibration of radius as a function of spectral type from \citet{2000asqu.book.....C} and the parallax of the host star. The masses for the exoplanet host stars and masses and radii for the possible secondary companions were obtained from \citet{2000asqu.book.....C}, which in turn were based on values derived from \citet{1981A&AS...46..193H}, who used observations of binary stars in order to create empirical relationships between various stellar parameters as a function of luminosity class.

Tables \ref{host_obs_props}, \ref{poss_sec_params}, and \ref{secondary_diams} present the results of these calculations. For each exoplanet host star studied, the observed values required for the calculations described above are listed in Table \ref{host_obs_props}, while Table \ref{poss_sec_params} shows the calculated $\Delta K$ and $\alpha$ for each type of possible secondary star. Table \ref{secondary_diams} includes the calculated angular diameters of the potential secondary companions for each system. It should be noted that if the companion star is a pre-main-sequence star, the resulting $\Delta K$ becomes smaller due to the star's increased brightness prior to hydrogen fusion and therefore has a higher probability of being detected.

The resulting values for $\theta$, $\Delta K$, and $\alpha$ were then used to plot the visibility curves for both a single star with the host star's measured angular diameter and for a binary system with the calculated parameters. The projected position angle of a binary star vector separation onto the interferometric baseline is, of course, unknown and is here assumed to be 0$^\circ$ to explore the effects of the maximum separation exhibited by the secondary. As this angle approaches 90$^{\circ}$, the modulation to the visibility curve diminishes because the binary becomes unresolved at 90$^{\circ}$.

To estimate the detection sensitivity, the largest difference between the visibility curves for a single star and for a binary system with the parameters listed in Table \ref{poss_sec_params} was calculated. This quantity, $\Delta V_{\rm max}$, then represented the maximum deviation of the binary visibility curve from the single-star curve.

The lower limit to rule out stellar companions was selected to be 2$\sigma_{\rm res}$, where $\sigma_{\rm res}$ is the standard deviation of the residuals to the diameter fit; i.e., if $\Delta V_{\rm max}$ $\geq$ $2 \sigma_{\rm res}$ for a given secondary component, that particular spectral type can be eliminated as a possible stellar companion. If $\Delta V_{\rm max}$ $\leq$ $2 \sigma_{\rm res}$, the effects of the companion would not be clearly seen in the visibility curve, and that spectral type cannot be ruled out. For each exoplanet host star, Table \ref{compare_resids} lists the observed $\sigma_{\rm res}$ and the predicted $\Delta V_{\rm max}$ for each secondary type considered, and the final column indicates the cutoff point for the non-detection of a stellar companion. For example, if ``K5~V'' is listed in the last column, the spectral types more massive than a K5~V could be eliminated from consideration but stars of type K5~V and later are still possible companions. A dash in this column indicates all companion spectral types can be ruled out.

\begin{figure}[!h]
  \centering \includegraphics[angle=90,width=1.0\textwidth]
  {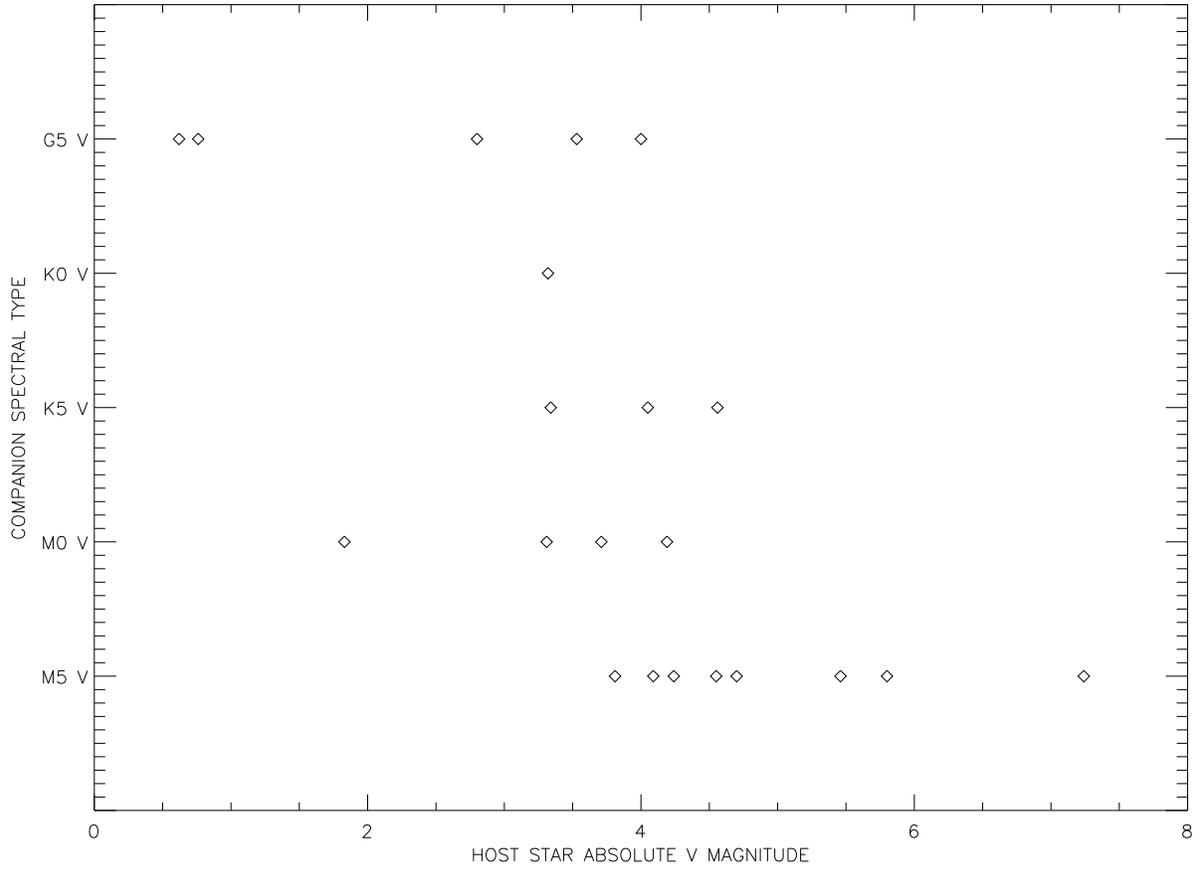}\\
  \caption{Companion check: Eliminating companion spectral types. The x-axis represents the absolute magnitude for the observed exoplanet host stars and the y-axis represents the lowest-mass secondary that is still a possibility for each observed star.}
  \label{bv_comptype}
\end{figure}

Using the data from Table \ref{compare_resids}, Figure \ref{bv_comptype} was created to demonstrate the sensitivity of the interferometric observations to stellar companions. For each exoplanet host star observed, the absolute $V$-band magnitude was found from \emph{The Hipparcos and Tycho Catalogues} \citep{1997A&A...323L..49P}. Then the night with the lowest $\sigma_{\rm res}$ was chosen to be the best case scenario when determining which secondary stars could be eliminated from consideration.

The difficulty of ruling out the more massive companion types for the intrinsically brighter stars is expected, as the $\Delta K$ will already be large even for brighter companions and beyond the scope of the CHARA Array. Host stars that are less massive are fainter, and the $\Delta K$ lies more within the sensitivity limit of the CHARA Array.

Two stars showed systematics in their visibility measurements that could indicate an unseen stellar companion in some datasets. The first star is $\upsilon$~Andromedae ($\upsilon$~And, HD~9826, F8~V). Its first planetary candidate was announced by \citet{1999ApJ...526..916B} and two more planets were discovered two years later \citep{1999ApJ...526..916B}. An M4.5~V stellar companion was found accompanying $\upsilon$~And at a distance of $\sim$750~AU from the central star \citep{2002ApJ...572L..79L}. This star would not have affected our search for more close-in stellar companions, as the angular distance from the host star is 55$^{\prime \prime}$ and is well out of the field of view of the CHARA Array.

$\upsilon$~And was part of two intensive observing campaigns using the CHARA Array in 2005 August and 2007 September and the data cannot be fit with a simple limb-darkened disk for two of the nine nights of data. In one dataset (2005/08/04), the target's visibilities briefly became higher than the calibrator's visibilities, and in the other dataset (2005/08/10), the target's and calibrator's visibilities separated over time. These patterns indicate one of the stars is changing with respect to the other. Two different calibrators were used for these two nights of data, and other data obtained using the same calibrators show no systematics in the visibilities. Therefore, we do not claim a stellar companion to $\upsilon$~And at this time.

The second star to show oddities in its visibility measurements was $\rho$~Coronae Borealis ($\rho$~CrB, HD~143761, G0~V). A planetary companion to $\rho$~CrB was announced by \citet{1997ApJ...483L.111N} before a later study derived a face-on orbit with an M dwarf, not a planet \citep{2001ApJ...548L..61G}. This claim was then refuted by \citet{2005AJ....129..402B}, who used high-dispersion infrared spectroscopy to determine if they could detect any flux from an M dwarf companion, as it would lie within the sensitivity limits of their instrument. No such flux was detected and they concluded the companion was planetary in nature.

The controversy surrounding this system made it an interesting target to observe using the CHARA Array. We observed $\rho$~CrB for three nights using three different calibrators and found that the data for two of the four nights exhibited behavior inconsistent with a single star. The two nights in question are 2005/06/29 and 2005/07/03, and because those observations were taken using the same calibrator and the data taken using other calibrators fall in line with a single star, it is likely that it was the calibrator and not $\rho$~CrB that was varying. There are currently plans to image both $\upsilon$~And and $\rho$~CrB using the Michigan Infrared Combiner \citep{2004SPIE.5491.1370M} on the CHARA Array, which may help to clarify the situation for both stars.


\section{Separated Fringe Packets}
The second method to check for unseen low-mass stellar companions is by searching for separated fringe packets (SFPs). When a star has a wide companion ($\sim$10 to 100 mas), two fringe packets - one from each star - may be observed if the baseline orientation is favorable; i.e., if the projected baseline angle is approximately parallel to the position angle of the binary and both fringe packets are within the data collection scan window. However, if the two stars have a small angular separation or the position angle is perpendicular to the projected baseline angle, the two fringe packets will overlay each other and appear as one fringe packet.

\begin{figure}[!h]
  \centering \includegraphics[width=1.0\textwidth]
  {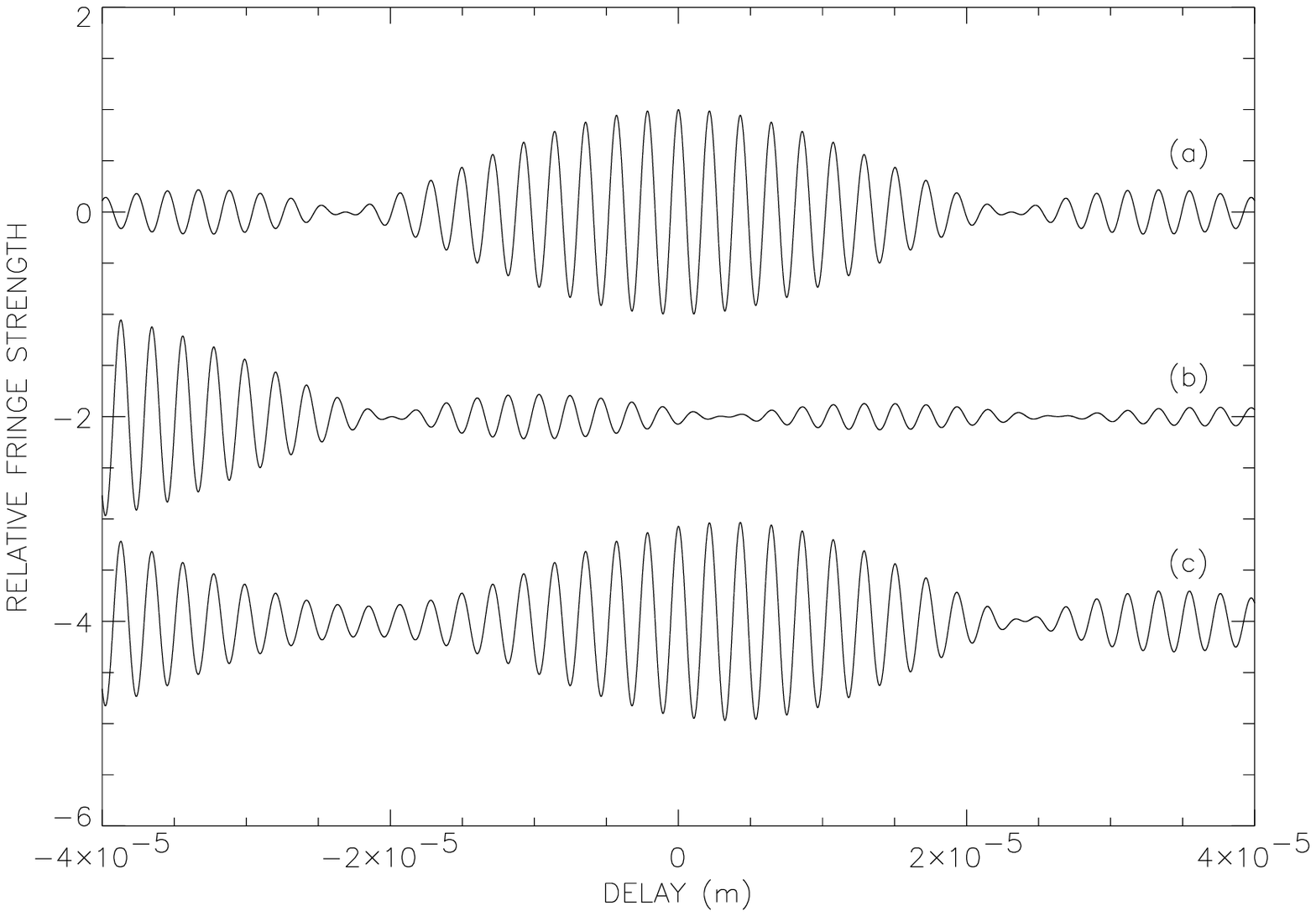}\\
  \caption{Example SFP for a $\Delta K$=0 binary system: (a) shows the primary star's fringe, (b) shows the secondary star's fringe, and (c) shows the combination of the two.}
  \label{exp_sfp_deltam0}
\end{figure}

\begin{figure}[!h]
  \centering \includegraphics[width=1.0\textwidth]
  {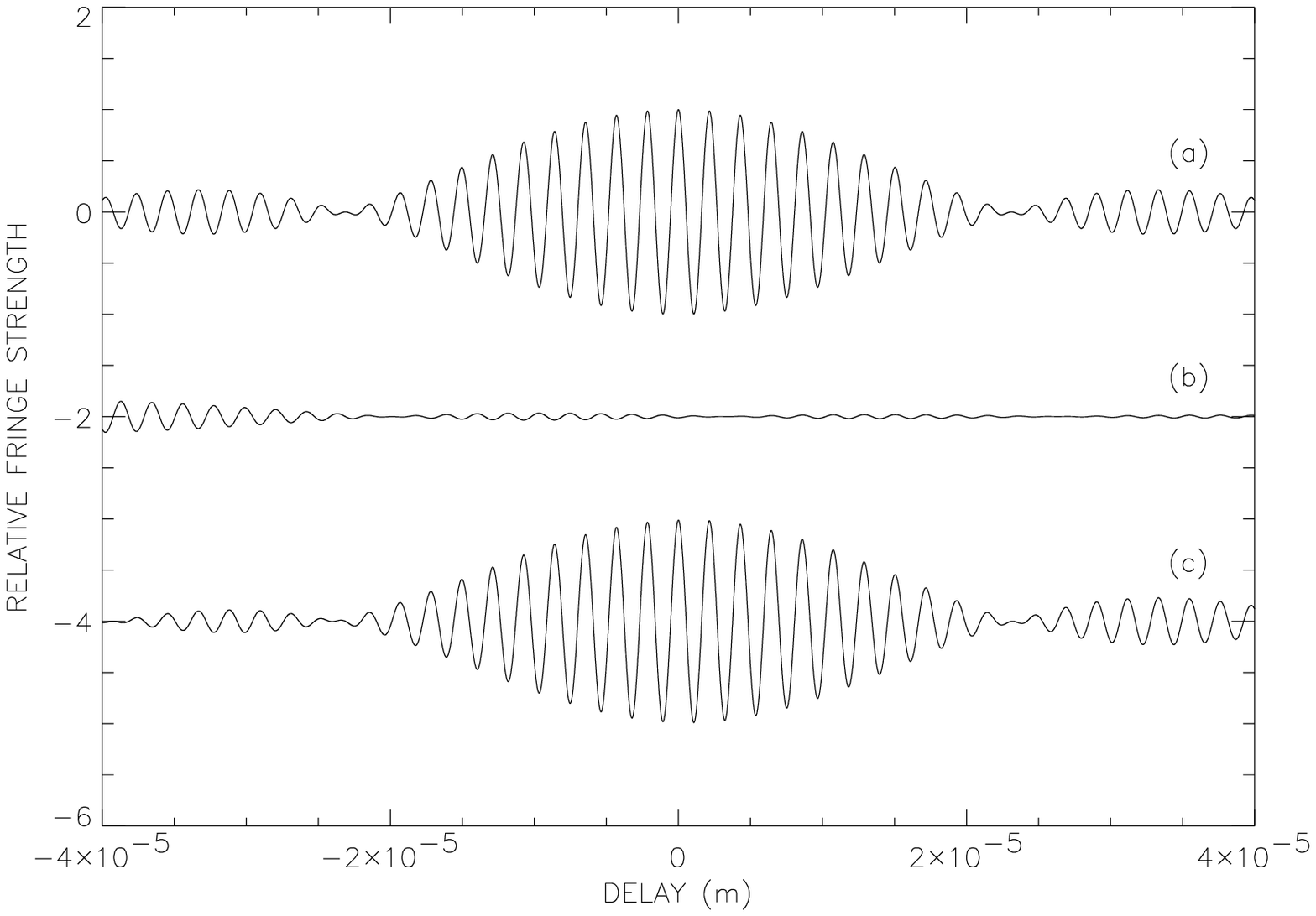}\\
  \caption{Example SFP for a $\Delta K$=2.5 binary system: (a) shows the primary star's fringe, (b) shows the secondary star's fringe, and (c) shows the combination of the two.}
  \label{exp_sfp_deltam2}
\end{figure}

For example, if a binary system with a $\Delta K$=0 has a separation of more than $\sim$10~mas and is in the optimal orientation as described above, SFPs may be visible, as can be seen in Figure \ref{exp_sfp_deltam0}. If the same system has a $\Delta K \sim$2.5 or is not in the optimal orientation, no secondary fringe would be observed, as is shown in Figure \ref{exp_sfp_deltam2}. The baseline used in the observations also plays a role in whether or not SFPs will be detected, as the separation of the fringes depends on the baseline length. Therefore, a system appearing as an SFP on the long baseline of S1-E1 will not be an SFP on the shortest baseline, S1-S2.

The detection of SFPs also depends partly on whether both fringe packets lie within the scan window. The width of the scan window depends on the baseline, wavelength used, and the frequency of the observations. For an average observation in the $K$-band using a 100-m baseline, the scan window will cover $\sim$300 mas while at a 300-m baseline, the scan window width is $\sim$100 mas. If the SFP is wider than the scan window width, data on the second fringe cannot be collected.

For completeness, all the stars observed in this project were checked for SFPs, whether or not the calculated separation of the secondary star would indicate the possibility of separated fringes. Each of the $\sim$200 scans in every dataset consists of a sampling of the zero-path-length delay space where fringes are located. For each individual scan, the strongest fringe was located and an envelope was fit to the fringe. The fringe envelope was obtained using a Hilbert tranform, achieved by taking the Fourier transform of the fringe scan, setting the negative frequencies to zero, and taking the modulus of the inverse Fourier transform \citep{1959prop.book.....B}.

\begin{figure}[!h]
  \centering \includegraphics[width=1.0\textwidth]
  {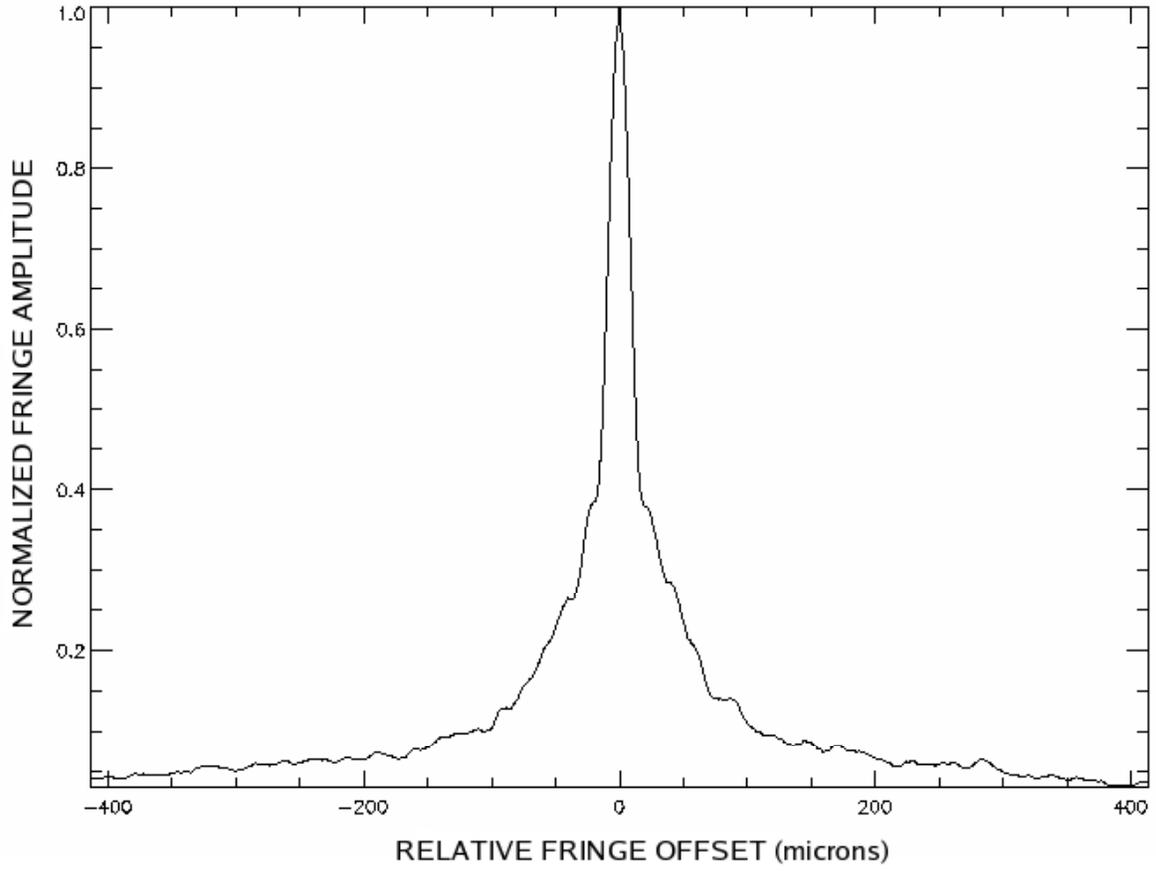}\\
  \caption{Example fringe envelope for a single star. The data for this star shows no indication of a secondary peak in the fringe envelope. (Data for HD~16141.)}
  \label{exp_frgenv}
\end{figure}

\begin{figure}[!h]
  \centering \includegraphics[width=1.0\textwidth]
  {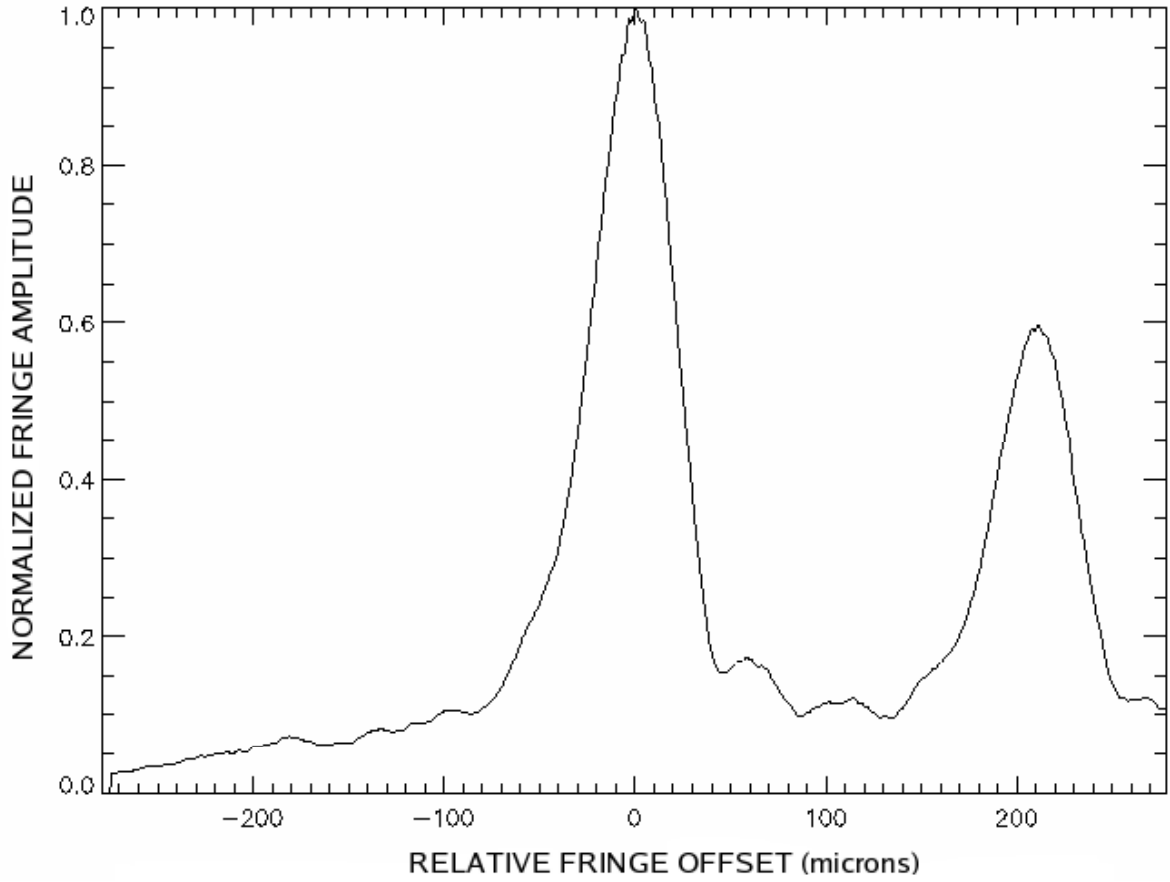}\\
  \caption{Example fringe envelope for a separated fringe packet binary. The secondary peak represents the stellar companion. (Data for HD~3196. Image courtesy of Deepak Raghavan.)}
  \label{exp_frgenv_sfp}
\end{figure}

Then the peak of the primary fringe envelope for each of the data scans was located and shifted so that the fringe envelopes for all the scans overlaid each other and the fringe amplitudes were added together. This ``shift and add'' approach made it possible to view multiple fringe envelopes at once to see if there was any indication of a SFP. The result was a plot of the weighted mean fringe envelope. Figures~\ref{exp_frgenv} and \ref{exp_frgenv_sfp} shows an examples of the fringe envelopes for a single star and binary system, respectively.

If there were three or fewer bracketed observations in the dataset, only one observation from each of the calibrator and object stars was inspected for SFPs. Otherwise the first and last observations in a night's dataset were inspected from SFPs for both the object and the calibrator in order to maximize changes in the stars' position angles with respect to the baseline over time. No SFP binaries were discovered for the exoplanet host stars or their calibrators. 

\section{Conclusion}
In an effort to cull out any possible binary star systems in the exoplanet sample, we inspected the exoplanet host stars using the CHARA Array using two methods. The first involved characterizing the residuals of the diameter fit to calibrated visibilities. If the observed residuals were more than twice the predicted variations for a given binary system, the secondary spectral type in question could be ruled out. No stellar companions were detected but certain secondary spectral types were eliminated for each exoplanet host star.

We also inspected the data for secondary fringe packets, which could be present if a secondary star had the proper orientation to the baseline used for the observations. No secondary fringe packet binaries were found, further reducing the possibility that these exoplanet systems are binary star systems instead.

\acknowledgements

Many thanks to P.J. Goldfinger and Chris Farrington for their invaluable assistance in obtaining the data used here. The CHARA Array is funded by the National Science Foundation through NSF grants AST-0307562 and AST-0606958 and by Georgia State University through the College of Arts and Sciences. This research has made use of the SIMBAD literature database, operated at CDS, Strasbourg, France, and of NASA's Astrophysics Data System. This publication makes use of data products from the Two Micron All Sky Survey, which is a joint project of the University of Massachusetts and the Infrared Processing and Analysis Center/California Institute of Technology, funded by the National Aeronautics and Space Administration and the National Science Foundation.

\clearpage

\begin{deluxetable}{cccccc}
\tablewidth{0pc}
\tabletypesize{\scriptsize}
\tablecaption{Interferometric Observations.\label{observations}}

\tablehead{ \colhead{Target} & \colhead{Other} & \colhead{Calibrator} & \colhead{Baseline} & \colhead{Date} & \colhead{Number of} \\
\colhead{HD} & \colhead{Name} & \colhead{HD} & \colhead{(length)} & \colhead{(UT)} & \colhead{Observations} \\ }
\startdata
3651 & 54 Psc & 4568 & S1-E1 (331 m) & 2005/10/24 & 6 \\
\hline
9826 & $\upsilon$ And & 6920 & W1-W2 (108 m) & 2005/08/04 & 11 \\
     &                &      &               & 2005/08/08 & 14 \\
     &                &      &               & 2005/08/14 & 10 \\
     &                &      & W1-S2 (249 m) & 2007/09/05 & 15 \\
     &                &      & S1-E1 (331 m) & 2004/01/14 & 13 \\
     &                &      &               & 2004/01/15 & 5  \\
     &                & 8671 & W1-W2 (108 m) & 2005/08/10 & 16 \\
     &                &      &               & 2005/08/18 & 20 \\
     &                &      &               & 2005/08/19 & 18 \\
     &                & 9712 & W1-S2 (249 m) & 2007/09/06 & 10 \\
     &                &      &               & 2007/09/12 & 8 \\

\hline
11964 & $\ldots$ & 13456 & S1-E1 (331 m) & 2007/09/14 & 6 \\
      &          &       &               & 2007/09/15 & 5 \\
\hline
12661 & $\ldots$ & 12846 & W1-W2 (108 m) & 2006/10/20 & 7 \\
      &          &       & W1-S1 (279 m) & 2005/12/16 & 5 \\
\hline
16141 &          &       & S1-E1 (331 m) & 2005/12/12 & 8  \\
\hline
19994 & 94 Cet & 19411 & S1-E1 (331 m) & 2005/10/27 & 6  \\
      &        &       &               & 2005/12/10 & 6  \\
\hline
20367 & $\ldots$ & 21864 & S1-E1 (331 m) & 2005/12/12 & 5  \\
\hline
23596 & $\ldots$ & 22521 & S1-E1 (331 m) & 2007/09/11 & 7 \\
      &          &       &               & 2007/09/14 & 5 \\
\hline
38529 & $\ldots$ & 43318 & W1-S1 (279 m) & 2005/12/14 & 5  \\
      &          &       & S1-E1 (331 m) & 2005/12/06 & 8  \\
\hline
59686 & $\ldots$ & 61630 & S1-E1 (331 m) & 2005/12/06 & 8  \\
      &          &       &               & 2005/12/16 & 8  \\
      &          &       &               & 2007/04/02 & 9  \\
\hline
75732 & 55 Cnc & 72779 & S1-E1 (331 m) & 2007/03/26 & 5  \\
      &        &       &               & 2007/03/30 & 6  \\
\hline
104985 & $\ldots$ & 97619 & W1-W2 (108 m) & 2006/05/17 & 10 \\
       &          &       & E1-W1 (314 m) & 2007/04/26 & 7 \\
\hline
117176 & 70 Vir & 121107 & W1-W2 (108 m) & 2006/05/13 & 10 \\
       &        &        & S1-E1 (331 m) & 2006/05/20 & 5  \\
       &        &        &               & 2007/04/02 & 6  \\
\hline
120136 & $\tau$ Boo & 121107 & W1-W2 (108 m) & 2006/05/14 & 11 \\
       &            &        & E2-W2 (156 m) & 2005/05/12 & 9 \\
       &            &        & S1-E1 (331 m) & 2007/02/05 & 10 \\
       &            &        &               & 2007/03/26 & 5  \\
       &            &        &               & 2007/03/30 & 8  \\
\hline
143761 & $\rho$ CrB & 143687 & W1-W2 (108 m) & 2006/05/12 & 8  \\
       &            & 143393 & E2-W2 (156 m) & 2005/06/29 & 6  \\
       &            &        &               & 2005/07/03 & 7  \\
       &            & 146025 & W1-W2 (108 m) & 2006/05/12 & 5  \\
\hline
177830 & $\ldots$ & 176377 & W1-W2 (108 m) & 2006/08/08 & 7  \\
       &          &        & S1-E1 (331 m) & 2006/08/13 & 6  \\
\hline
186427 & 16 Cyg B & 184960 & S1-E1 (331 m) & 2006/08/13 & 6  \\
       &          &        &               & 2007/09/12 & 6  \\
\hline
190228 & $\ldots$ & 190470 & W1-W2 (108 m) & 2005/08/19 & 5 \\
       &          &        & E2-W2 (156 m) & 2005/07/01 & 6 \\
       &          &        & S1-E1 (331 m) & 2006/08/14 & 8 \\
\hline
190360 & $\ldots$ & 189108 & W1-W2 (108 m) & 2005/08/11 & 10 \\
       &          &        & S1-E1 (331 m) & 2006/08/11 & 9  \\
\hline
195019 & $\ldots$ & 194012 & W1-W2 (108 m) & 2005/08/11 & 5  \\
       &          &        &               & 2006/08/06 & 6  \\
       &          &        & S1-E1 (331 m) & 2005/10/23 & 10 \\
\hline
196885 & $\ldots$ & 194012 & W1-W2 (108 m) & 2006/08/07 & 10 \\
       &          &        & E2-W2 (156 m) & 2005/10/29 & 5  \\
       &          &        & S1-E1 (331 m) & 2006/08/14 & 5  \\
\hline
217014 & 51 Peg A & 218261 & W1-W2 (108 m) & 2005/08/12 & 14 \\
       &          &        & S1-E1 (331 m) & 2006/08/12 & 7  \\
\enddata
\tablecomments{The three arms of the Array are denoted by their cardinal directions: ``S'' is south, ``E'' is east, and ``W'' is west. Each arm bears two telescopes, numbered ``1'' for the telescope farthest from the beam combining laboratory and ``2'' for the telescope closer to the lab.}
\end{deluxetable}

\clearpage

\begin{deluxetable}{ccccccccccccccc}
\rotate
\tablewidth{1.15\textwidth}
\tabletypesize{\scriptsize}
\tablecaption{Exoplanet Host Star Observed Parameters \label{host_obs_props}}

\tablehead{\multicolumn{3}{c}{Observed} & \multicolumn{1}{c}{ } &\multicolumn{7}{c}{Planetary Parameters} & \multicolumn{ 1}{c}{ } & \multicolumn{ 3}{c}{Calculated Values} \\
\cline{1-3} \cline{5-11} \cline{13-15}  \\ 

\colhead{ } &  \colhead{$K$} & \colhead{$\pi$} & \colhead{ } & \colhead{$M_{\rm star}$} &  \colhead{$P$} & \colhead{ } & \colhead{$K_{\rm 1}$} &  \colhead{$a_{\rm p}$} & \colhead{$M_{\rm p} \sin i$} & \colhead{ } & \colhead{ } & \colhead{ } & \colhead{$a_{\rm star} \sin i$} & \colhead{f($M_{\rm star}$)}  \\

\colhead{HD} & \colhead{(mag)} & \colhead{(mas)} &  \colhead{ } & \colhead{$M_\odot$} & \colhead{(d)} & \colhead{$e$} &  \colhead{(m s$^{-1}$)} & \colhead{(AU)} & \colhead{($M_{\rm Jup}$)} & \colhead{Ref} & \colhead{ } & \colhead{(m-M)} & \colhead{(AU)} &  \colhead{($M_\odot$)} \\ }
\startdata
3651 & 4.0 & 90.0 & & 0.79 & 62.2 & 0.63 & 15.9 & 0.28 & 0.20 & F03 & & 0.2 & 7.06E-05 & 1.21E-11 \\
\hline
9826 & 2.9 & 74.3 & & 1.3 & 4.6 & 0.03 & 73.0 & 0.06 & 0.71 & B99 & & 0.6 & 3.10E-05 & 1.86E-10 \\
\hline
11964 & 4.5 & 29.4 & & 1.12 & 2110 & 0.06 & 9.0 & 3.34 & 0.61 & B06 & & 2.7 & 1.74E-03 & 1.58E-10 \\
\hline
16141 & 5.3 & 27.9 & & 1.01 & 75.8 & 0.28 & 10.8 & 0.35 & 0.22 & M00 & & 2.8 & 7.23E-05 & 8.75E-12 \\
\hline
19994 & 3.7 & 44.7 & & 1.34 & 535.7 & 0.30 & 36.2 & 1.42 & 1.68 & M04 & & 1.7 & 1.70E-03 & 2.29E-09 \\
\hline
20367 & 5.0 & 36.9 & & 1.17 & 469.5 & 0.32 & 29.0 & 1.25 & 1.17 & B06 & & 2.2 & 1.19E-03 & 1.01E-09 \\
\hline
23596 & 5.9 & 19.2 & & 1.3 & 1565 & 0.29 & 124 & 2.88 & 8.1 & P03 & & 3.6 & 1.71E-02 & 2.71E-07 \\
\hline
38529 & 4.2 & 23.6 & & 1.4 & 14.3 & 0.27 & 53.8 & 0.13 & 0.77 & F01 & & 3.1 & 6.81E-05 & 2.06E-10 \\
\hline
59686 & 2.9 & 10.8 & & 1.15$^\ddagger$ & 303 & 0 & 10.0$^\dagger$ & $\ldots$ & 6.5 & M03 & & 4.8 & 2.79E-04 & 3.14E-11 \\
\hline
75732 & 4.0 & 79.8 & & 0.95 & 14.7 & 0.02 & 72.2 & 0.12 & 0.84 & M02 & & 0.5 & 9.73E-05 & 5.71E-10 \\
\hline
104985 & 3.3 & 9.8 & & 1.6 & 198.2 & 0.03 & 161 & 0.78 & 6.3 & S03 & & 5.0 & 2.93E-03 & 8.56E-08 \\
\hline
117176 & 3.5 & 55.2 & & 0.92 & 116.7 & 0.40 & 318 & 0.43 & 6.6 & MB96 & & 1.3 & 3.13E-03 & 2.99E-07 \\
\hline
120136 & 3.5 & 64.1 & & 1.2 & 3.3 & 0.02 & 469 & 0.05 & 3.87 & B97 & & 1.0 & 1.43E-04 & 3.54E-08 \\
\hline
143761 & 3.9 & 57.4 & & 1.0 & 39.6 & 0.03 & 67.4 & 0.23 & 1.1 & N97 & & 1.2 & 2.46E-04 & 1.26E-09 \\
\hline
177830 & 4.8 & 16.9 & & 1.15 & 391.6 & 0.41 & 34 & 1.10 & 1.22 & V00 & & 3.9 & 1.12E-03 & 1.21E-09 \\
\hline
186427 & 4.7 & 46.7 & & 1.0 & 800.8 & 0.63 & 43.9 & 1.6 & 1.5 & C97 & & 1.7 & 2.51E-03 & 3.29E-09 \\
\hline
190228 & 5.4 & 16.1 & & 0.83 & 1146 & 0.50 & 91 & 2.02 & 3.58 & P03 & & 4.0 & 8.30E-03 & 5.81E-08 \\
\hline
190360 & 4.1 & 62.9 & & 0.96 & 3902 & 0.48 & 20 & 4.8 & 1.33 & N03 & & 1.0 & 6.29E-03 & 2.18E-09 \\
\hline
195019 & 5.3 & 26.8 & & 0.98 & 18.3 & 0.03 & 275.3 & $\ldots$ & 3.51 & F99 & & 2.9 & 4.63E-04 & 3.95E-08 \\
\hline
196885 & 5.1 & 30.3 & & 1.27 & 386 & 0.3 & 10.0$^\dagger$ & 1.12 & 1.84 & EE & & 2.6 & 3.38E-04 & 3.47E-11 \\
\hline
217014 & 3.9 & 65.1 & & 1.0 & 4.2 & 0.01 & 55.9 & 0.05 & 0.45 & M97 & & 0.9 & 2.17E-05 & 7.65E-11 \\
\enddata
\tablecomments{$^\dagger$HD~59686 and HD~196885 had no $K_1$ given in the literature, so a $K_1$ of 10.0 m~s$^{-1}$ was assigned for the calculations. \\
$^\ddagger$HD 59686's mass was obtained from \citet{2000asqu.book.....C}. \\
All $K$ magnitudes are from the \emph{2MASS All-Sky Catalog of Point Sources} \citep{2006AJ....131.1163S} and $\pi$ values are from \emph{The Hipparcos Catalogue} \citep{1997A&A...323L..49P}. \\
References: B97: \citet{1997ApJ...474L.115B}; B99: \citet{1999ApJ...526..916B}; B06: \citet{2006ApJ...646..505B}; C97: \citet{1997ApJ...483..457C}; EE: Extrasolar Planets Encyclopaedia; F99: \citet{1999PASP..111...50F}; F01: \citet{2001ApJ...551.1107F}; F03: \citet{2003ApJ...590.1081F}; M97: \citet{1997ApJ...481..926M}; M00: \citet{2000ApJ...536L..43M}; M02: \citet{2002ApJ...581.1375M}; M03: \citet{2003AAS...203.1703M}; M04: \citet{2004A&A...415..391M}; MB96: \citet{1996ApJ...464L.147M}; N97: \citet{1997ApJ...483L.111N}; N03: \citet{2003A&A...410.1051N}; P03: \citet{2003A&A...410.1039P}; S03: \citet{2003ApJ...597L.157S}; V00: \citet{2000ApJ...536..902V}.
}
\end{deluxetable}

\clearpage

\begin{deluxetable}{cccccccccccccccc}
\rotate
\tablewidth{1.03\textwidth}
\tabletypesize{\scriptsize}
\tablecaption{Predicted Parameters for Secondary Stars of Various Spectral Types \label{poss_sec_params}}

\tablehead{\multicolumn{1}{c}{ } & \multicolumn{1}{c}{ } & \multicolumn{2}{c}{G5 V} & \multicolumn{1}{c}{ } & \multicolumn{2}{c}{K0 V} & \multicolumn{1}{c}{ } & \multicolumn{2}{c}{K5 V} & \multicolumn{1}{c}{ } & \multicolumn{2}{c}{M0 V} & \multicolumn{1}{c}{ } & \multicolumn{2}{c}{M5 V} \\
\cline{3-4} \cline{6-7} \cline{9-10} \cline{12-13} \cline{15-16}  \\ 
\colhead{Host} & \colhead{ } & \colhead{$\Delta K$} & \colhead{$\alpha$} & \colhead{ } & \colhead{$\Delta K$} & \colhead{$\alpha$} & \colhead{ } & \colhead{$\Delta K$} & \colhead{$\alpha$} & \colhead{ } & \colhead{$\Delta K$} & \colhead{$\alpha$} & \colhead{ } & \colhead{$\Delta K$} & \colhead{$\alpha$}  \\
\colhead{HD} & \colhead{ } & \colhead{(mag)} & \colhead{(mas)} &  \colhead{ } & \colhead{(mag)} & \colhead{(mas)} & \colhead{ } & \colhead{(mag)} & \colhead{(mas)} & \colhead{ } & \colhead{(mag)} & \colhead{(mas)} & \colhead{ } & \colhead{(mag)} & \colhead{(mas)} \\ }
\startdata
3651 &   &  0.27 & 1.15 &   & 0.13 & 1.12 &   & 0.73 & 1.09 &   & 1.38 & 1.05 &   & 2.36 & 0.97 \\
9826 &   & 1.25 & 0.22 &   & 1.65 & 0.22 &   & 2.25 & 0.21 &   & 2.90 & 0.21 &   & 3.88 & 0.20 \\
11964 &   & 1.66 & 12.83 &   & 2.06 & 12.55 &   & 2.66 & 12.28 &   & 3.31 & 11.90 &   & 4.29 & 11.12 \\
16141 &   & 0.98 & 1.37 &   & 1.38 & 1.34 &   & 1.98 & 1.31 &   & 2.63 & 1.27 &   & 3.61 & 1.18 \\
19994 &   & 1.55 & 5.32 &   & 1.95 & 5.22 &   & 2.55 & 5.12 &   & 3.20 & 4.98 &   & 4.18 & 4.69 \\
20367 &   & 0.66 & 4.75 &   & 1.06 & 4.65 &   & 1.66 & 4.55 &   & 2.31 & 4.42 &   & 3.29 & 4.13 \\
23596 &   & 1.20 & 10.81 &   & 1.60 & 10.60 &   & 2.20 & 10.39 &   & 2.85 & 10.10 &   & 3.83 & 9.51 \\
38529 &   & 2.44 & 0.48 &   & 2.84 & 0.47 &   & 3.44 & 0.46 &   & 4.09 & 0.45 &   & 5.07 & 0.42 \\
59686 &   & 5.43 & 3.53 &   & 5.83 & 3.46 &   & 6.43 & 3.39 &   & 7.08 & 3.28 &   & 8.06 & 3.07 \\
75732 &   &  0.01 & 0.45 &   & 0.39 & 0.44 &   & 0.99 & 0.43 &   & 1.64 & 0.42 &   & 2.62 & 0.39 \\
104985 &   & 5.24 & 2.84 &   & 5.64 & 2.79 &   & 6.24 & 2.75 &   & 6.89 & 2.68 &   & 7.87 & 2.55 \\
117176 &   & 1.29 & 1.80 &   & 1.69 & 1.76 &   & 2.29 & 1.71 &   & 2.94 & 1.65 &   & 3.92 & 1.53 \\
120136 &   & 0.97 & 0.18 &   & 1.37 & 0.17 &   & 1.97 & 0.17 &   & 2.62 & 0.16 &   & 3.60 & 0.15 \\
143761 &   & 0.81 & 0.89 &   & 1.21 & 0.87 &   & 1.81 & 0.85 &   & 2.46 & 0.82 &   & 3.44 & 0.76 \\
177830 &   & 2.56 & 4.19 &   & 2.96 & 4.10 &   & 3.56 & 4.02 &   & 4.21 & 3.90 &   & 5.19 & 3.65 \\
186427 &   & 0.45 & 6.59 &   & 0.85 & 6.44 &   & 1.45 & 6.29 &   & 2.10 & 6.08 &   & 3.08 & 5.65 \\
190228 &   & 2.07 & 8.11 &   & 2.47 & 7.91 &   & 3.07 & 7.71 &   & 3.72 & 7.42 &   & 4.70 & 6.82 \\
190360 &   & 0.41 & 18.81 &   & 0.81 & 18.36 &   & 1.41 & 17.93 &   & 2.06 & 17.33 &   & 3.04 & 16.06 \\
195019 &   & 1.06 & 0.53 &   & 1.46 & 0.52 &   & 2.06 & 0.50 &   & 2.71 & 0.49 &   & 3.69 & 0.45 \\
196885 &   & 0.99 & 4.23 &   & 1.39 & 4.15 &   & 1.99 & 4.07 &   & 2.64 & 3.95 &   & 3.62 & 3.71 \\
217014 &   & 0.53 & 0.20 &   & 0.93 & 0.20 &   & 1.53 & 0.19 &   & 2.18 & 0.18 &   & 3.16 & 0.17 \\
\enddata
\tablecomments{Values for M$_K$ were obtained from \citet{2000asqu.book.....C}: M$_{K \rm{(G5 V)}} = 3.5$, M$_{K \rm{(K0 V)}} = 3.9$, M$_{K\rm{(K5 V)}} = 4.5$, M$_{K\rm{(M0 V)}} = 5.2$, and M$_{K\rm{(M5 V)}} = 6.1$.}
\end{deluxetable}

\clearpage

\begin{deluxetable}{ccccccc}
\tablewidth{0pc}
\tabletypesize{\scriptsize}
\tablecaption{Calculated Secondary Star Angular Diameters \label{secondary_diams}}

\tablehead{\multicolumn{1}{c}{Host Star} & \multicolumn{1}{c}{ } &\multicolumn{5}{c}{Angular Diameter (mas)}  \\
\colhead{HD} &  \colhead{ } &\colhead{G5 V} & \colhead{K0 V} & \colhead{K5 V} & \colhead{M0 V} &  \colhead{M5 V} \\ }
\startdata
3651 &   & 0.77 & 0.71 & 0.60 & 0.50 & 0.23 \\
9826 &   & 0.64 & 0.59 & 0.50 & 0.41 & 0.19 \\
11964 &   & 0.25 & 0.23 & 0.20 & 0.16 & 0.07 \\
16141 &   & 0.24 & 0.22 & 0.19 & 0.16 & 0.07 \\
19994 &   & 0.38 & 0.35 & 0.30 & 0.25 & 0.11 \\
20367 &   & 0.32 & 0.29 & 0.25 & 0.21 & 0.09 \\
23596 &   & 0.16 & 0.15 & 0.13 & 0.11 & 0.05 \\
38529 &   & 0.20 & 0.19 & 0.16 & 0.13 & 0.06 \\
59686 &   & 0.09 & 0.09 & 0.07 & 0.06 & 0.03 \\
75732 &   & 0.68 & 0.63 & 0.53 & 0.45 & 0.20 \\
104985 &   & 0.08 & 0.08 & 0.07 & 0.05 & 0.02 \\
117176 &   & 0.47 & 0.44 & 0.37 & 0.31 & 0.14 \\
120136 &   & 0.55 & 0.51 & 0.43 & 0.36 & 0.16 \\
143761 &   & 0.49 & 0.45 & 0.38 & 0.32 & 0.14 \\
177830 &   & 0.15 & 0.13 & 0.11 & 0.09 & 0.04 \\
186427 &   & 0.40 & 0.37 & 0.31 & 0.26 & 0.12 \\
190228 &   & 0.14 & 0.13 & 0.11 & 0.09 & 0.04 \\
190360 &   & 0.54 & 0.50 & 0.42 & 0.35 & 0.16 \\
195019 &   & 0.23 & 0.21 & 0.18 & 0.15 & 0.07 \\
196885 &   & 0.26 & 0.24 & 0.20 & 0.17 & 0.08 \\
217014 &   & 0.56 & 0.51 & 0.44 & 0.36 & 0.16 \\
\enddata
\tablecomments{The radii used in these calculations: G5~V~=~0.92~$R_\odot$; K0~V~=~0.85~$R_\odot$; K5~V~=~0.72~$R_\odot$; M0~V~=~0.60~$R_\odot$; M5~V~=~0.27~$R_\odot$. These were obtained from \citet{2000asqu.book.....C}.}
\end{deluxetable}

\clearpage

\begin{deluxetable}{ccccccccccc}
\tablewidth{0.85\textwidth}
\tabletypesize{\scriptsize}
\tablecaption{Companion Check: Comparing Visibility Residuals \label{compare_resids}}

\tablehead{\multicolumn{1}{c}{ } & \multicolumn{1}{c}{ } & \multicolumn{1}{c}{ } & \multicolumn{1}{c}{ } & \multicolumn{5}{c}{$\Delta V_{\rm max}$ of Secondary Spectral Type} & \multicolumn{1}{c}{ } & \multicolumn{1}{c}{Non Detection} \\ \cline{5-9}  \\ 
\colhead{HD} &  \colhead{Obs Date} & \colhead{$\sigma_{\rm res}$} & \colhead{ } & \colhead{G5 V} & \colhead{K0 V} & \colhead{K5 V} & \colhead{M0 V} &  \colhead{M5 V} & \colhead{ } & \colhead{Threshold} \\ }
\startdata
3651 & 2005/10/24 & 0.043 &   & 0.429 & 0.481 & 0.314 & 0.187 & 0.079 &   & M5 V \\
9826 & 2004/01/14 & 0.024 &   & 0.025 & 0.018 & 0.013 & 0.007 & 0.004 &   & G5 V \\
  & 2004/01/15 & 0.022 &   & 0.030 & 0.021 & 0.015 & 0.008 & 0.004 &   & G5 V \\
  & 2005/08/04 & 0.102 &   & 0.116 & 0.095 & 0.068 & 0.047 & 0.029 &   & G5 V \\
  & 2005/08/08 & 0.063 &   & 0.084 & 0.064 & 0.040 & 0.022 & 0.009 &   & G5 V \\
  & 2005/08/10 & 0.065 &   & 0.107 & 0.086 & 0.058 & 0.036 & 0.016 &   & G5 V \\
  & 2005/08/14 & 0.060 &   & 0.084 & 0.064 & 0.041 & 0.022 & 0.009 &   & G5 V \\
  & 2005/08/18 & 0.051 &   & 0.089 & 0.068 & 0.044 & 0.024 & 0.010 &   & G5 V \\
  & 2005/08/19 & 0.055 &   & 0.150 & 0.137 & 0.123 & 0.125 & 0.137 &   & - \\
  & 2007/09/05 & 0.061 &   & 0.031 & 0.022 & 0.015 & 0.008 & 0.004 &   & G5 V \\
  & 2007/09/06 & 0.062 &   & 0.034 & 0.024 & 0.016 & 0.005 & 0.005 &   & G5 V \\
  & 2007/09/12 & 0.052 &   & 0.042 & 0.030 & 0.020 & 0.011 & 0.005 &   & K0 V \\
11964 & 2005/12/16 & 0.033 &   & 0.204 & 0.145 & 0.086 & 0.050 & 0.020 &   & M0 V \\
  & 2006/10/20 & 0.070 &   & 0.222 & 0.160 & 0.095 & 0.054 & 0.022 &   & K5 V \\
16141 & 2005/12/12 & 0.250 &   & 0.334 & 0.243 & 0.148 & 0.083 & 0.033 &   & G5 V \\
19994 & 2005/10/27 & 0.034 &   & 0.229 & 0.164 & 0.098 & 0.055 & 0.022 &   & M0 V \\
  & 2005/12/10 & 0.037 &   & 0.229 & 0.164 & 0.098 & 0.055 & 0.022 &   & M0 V \\
20367 & 2005/12/12 & 0.042 &   & 0.402 & 0.300 & 0.190 & 0.114 & 0.050 &   & M5 V \\
23596 & 2007/09/11 & 0.050 &   & 0.278 & 0.207 & 0.127 & 0.071 & 0.030 &   & M0 V \\
  & 2007/09/14 & 0.032 &   & 0.270 & 0.201 & 0.123 & 0.069 & 0.029 &   & M5 V \\
38529 & 2005/12/06 & 0.055 &   & 0.081 & 0.057 & 0.033 & 0.018 & 0.007 &   & G5 V \\
  & 2005/12/14 & 0.155 &   & 0.066 & 0.046 & 0.027 & 0.015 & 0.006 &   & G5 V \\
59686 & 2005/12/06 & 0.007 &   & 0.007 & 0.005 & 0.003 & 0.002 & 0.001 &   & G5 V \\
  & 2005/12/16 & 0.038 &   & 0.008 & 0.005 & 0.003 & 0.002 & 0.001 &   & G5 V \\
  & 2007/04/02 & 0.011 &   & 0.007 & 0.005 & 0.003 & 0.002 & 0.001 &   & G5 V \\
75732 & 2007/03/26 & 0.018 &   & 0.101 & 0.109 & 0.103 & 0.077 & 0.037 &   & - \\
  & 2006/03/30 & 0.053 &   & 0.201 & 0.187 & 0.149 & 0.101 & 0.046 &   & M0 V \\
104985 & 2006/05/17 & 0.027 &   & 0.009 & 0.006 & 0.004 & 0.002 & 0.001 &   & G5 V \\
  & 2007/04/26 & 0.028 &   & 0.009 & 0.006 & 0.004 & 0.002 & 0.001 &   & G5 V \\
117176 & 2006/05/13 & 0.052 &   & 0.277 & 0.200 & 0.120 & 0.067 & 0.026 &   & M0 V \\
  & 2006/05/20 & 0.019 &   & 0.277 & 0.203 & 0.124 & 0.071 & 0.028 &   & M0 V \\
  & 2007/04/02 & 0.094 &   & 0.277 & 0.203 & 0.124 & 0.070 & 0.028 &   & K5 V \\
120136 & 2005/05/12 & 0.071 &   & 0.018 & 0.017 & 0.011 & 0.008 & 0.005 &   & G5 V \\
  & 2006/05/14 & 0.026 &   & 0.040 & 0.033 & 0.021 & 0.014 & 0.007 &   & G5 V \\
  & 2007/02/05 & 0.020 &   & 0.001 & 0.004 & 0.004 & 0.004 & 0.003 &   & G5 V \\
  & 2007/03/26 & 0.071 &   & 0.002 & 0.002 & 0.002 & 0.003 & 0.003 &   & G5 V \\
  & 2007/03/30 & 0.034 &   & 0.001 & 0.005 & 0.004 & 0.004 & 0.003 &   & G5 V \\
143761 & 2005/06/29 & 0.049 &   & 0.250 & 0.178 & 0.107 & 0.061 & 0.026 &   & Mo V \\
  & 2005/07/03 & 0.078 &   & 0.251 & 0.176 & 0.103 & 0.057 & 0.023 &   & K5 V \\
  & 2006/05/12 & 0.035 &   & 0.251 & 0.177 & 0.105 & 0.059 & 0.025 &   & M0 V \\
  & 2006/05/12 & 0.092 &   & 0.250 & 0.176 & 0.103 & 0.056 & 0.023 &   & K0 V \\
177830 & 2006/08/08 & 0.040 &   & 0.101 & 0.071 & 0.041 & 0.023 & 0.009 &   & K0 V \\
  & 2006/08/13 & 0.095 &   & 0.096 & 0.068 & 0.040 & 0.022 & 0.008 &   & G5 V \\
186427 & 2006/08/13 & 0.072 &   & 0.478 & 0.358 & 0.225 & 0.130 & 0.058 &   & M0 V \\
  & 2007/09/12 & 0.048 &   & 0.496 & 0.371 & 0.233 & 0.134 & 0.059 &   & M5 V \\
190228 & 2005/07/01 & 0.076 &   & 0.148 & 0.106 & 0.062 & 0.034 & 0.014 &   & G5 V \\
  & 2005/08/19 & 0.044 &   & 0.159 & 0.113 & 0.067 & 0.037 & 0.015 &   & K5 V \\
  & 2006/08/14 & 0.042 &   & 0.141 & 0.102 & 0.060 & 0.032 & 0.014 &   & K5 V \\
190360 & 2005/08/11 & 0.045 &   & 0.560 & 0.421 & 0.266 & 0.156 & 0.067 &   & M5 V \\
  & 2006/08/11 & 0.044 &   & 0.517 & 0.394 & 0.251 & 0.147 & 0.063 &   & M5 V \\
195019 & 2005/08/11 & 0.049 &   & 0.192 & 0.149 & 0.093 & 0.054 & 0.021 &   & K5 V \\
  & 2005/10/23 & 0.086 &   & 0.284 & 0.212 & 0.130 & 0.075 & 0.030 &   & K5 V \\
  & 2006/08/06 & 0.070 &   & 0.255 & 0.191 & 0.117 & 0.068 & 0.027 &   & K5 V \\
196885 & 2005/10/29 & 0.053 &   & 0.366 & 0.269 & 0.164 & 0.094 & 0.039 &   & M0 V \\
  & 2006/08/07 & 0.037 &   & 0.365 & 0.268 & 0.164 & 0.093 & 0.039 &   & M5 V \\
  & 2006/08/14 & 0.055 &   & 0.340 & 0.252 & 0.155 & 0.088 & 0.036 &   & M0 V \\
217014 & 2005/08/12 & 0.047 &   & 0.126 & 0.100 & 0.066 & 0.040 & 0.018 &   & K5 V \\
  & 2006/08/12 & 0.033 &   & 0.012 & 0.010 & 0.004 & 0.0005 & 0.002 &   & G5 V \\
\enddata
\end{deluxetable}

\clearpage

\clearpage

\clearpage

\clearpage

\clearpage

\clearpage

\clearpage

\clearpage

\begin{thebibliography}{99}

\bibitem[Bender et al.(2005)]{2005AJ....129..402B} Bender, C., Simon, M., Prato, L., Mazeh, T., \& Zucker, S.\ 2005, \aj, 129, 402

\bibitem[Born \& Wolf(1959)]{1959prop.book.....B} Born, M., \& Wolf, E.\ 1959, London: Pergamon Press, 1959

\bibitem[Butler et al.(1997)]{1997ApJ...474L.115B} Butler, R.~P., Marcy, G.~W., Williams, E., Hauser, H., \& Shirts, P.\ 1997, \apjl, 474, L115 

\bibitem[Butler et al.(1999)]{1999ApJ...526..916B} Butler, R.~P., Marcy, G.~W., Fischer, D.~A., Brown, T.~M., Contos, A.~R., Korzennik, S.~G., Nisenson, P., \& Noyes, R.~W.\ 1999, \apj, 526, 916 

\bibitem[Butler et al.(2006)]{2006ApJ...646..505B} Butler, R.~P., et al.\ 2006, \apj, 646, 505 

\bibitem[Cochran et al.(1997)]{1997ApJ...483..457C} Cochran, W.~D., Hatzes, A.~P., Butler, R.~P., \& Marcy, G.~W.\ 1997, \apj, 483, 457 

\bibitem[Cox(2000)]{2000asqu.book.....C} Cox, A.~N.\ 2000, Allen's astrophysical quantities, 4th ed., ed. A.~N.~Cox (New York: AIP Press; Springer)

\bibitem[Fischer et al.(1999)]{1999PASP..111...50F} Fischer, D.~A., Marcy, G.~W., Butler, R.~P., Vogt, S.~S., \& Apps, K.\ 1999, \pasp, 111, 50

\bibitem[Fischer et al.(2001)]{2001ApJ...551.1107F} Fischer, D.~A., Marcy, G.~W., Butler, R.~P., Vogt, S.~S., Frink, S., \& Apps, K.\ 2001, \apj, 551, 1107 

\bibitem[Fischer et al.(2003)]{2003ApJ...590.1081F} Fischer, D.~A., Butler, R.~P., Marcy, G.~W., Vogt, S.~S., \& Henry, G.~W.\ 2003, \apj, 590, 1081 

\bibitem[Gatewood et al.(2001)]{2001ApJ...548L..61G} Gatewood, G., Han, I., \& Black, D.~C.\ 2001, \apjl, 548, L61 

\bibitem[Habets \& Heintze(1981)]{1981A&AS...46..193H} Habets, G.~M.~H.~J., \& Heintze, J.~R.~W.\ 1981, \aaps, 46, 193 

\bibitem[Imbert \& Pr{\'e}vot(1998)]{1998A&A...334L..37I} Imbert, M., \& Pr{\'e}vot, L.\ 1998, \aap, 334, L37 

\bibitem[Lowrance et al.(2002)]{2002ApJ...572L..79L} Lowrance, P.~J., Kirkpatrick, J.~D., \& Beichman, C.~A.\ 2002, \apjl, 572, L79 

\bibitem[Marcy \& Butler(1996)]{1996ApJ...464L.147M} Marcy, G.~W., \& Butler, R.~P.\ 1996, \apjl, 464, L147

\bibitem[Marcy, Butler, \& Vogt(2000)]{2000ApJ...536L..43M} Marcy, G.~W., Butler, R.~P., \& Vogt, S.~S.\ 2000, \apjl, 536, L43 

\bibitem[Marcy et al.(1997)]{1997ApJ...481..926M} Marcy, G.~W., Butler, R.~P., Williams, E., Bildsten, L., Graham, J.~R., Ghez, A.~M., \& Jernigan, J.~G.\ 1997, \apj, 481, 926

\bibitem[Marcy et al.(2002)]{2002ApJ...581.1375M} Marcy, G.~W., Butler, R.~P., Fischer, D.~A., Laughlin, G., Vogt, S.~S., Henry, G.~W., \& Pourbaix, D.\ 2002, \apj, 581, 1375 

\bibitem[Mayor et al.(2004)]{2004A&A...415..391M} Mayor, M., Udry, S., Naef, D., Pepe, F., Queloz, D., Santos, N.~C., \& Burnet, M.\ 2004, \aap, 415, 391 

\bibitem[McAlister et al.(2005)]{2005ApJ...628..439M} McAlister, H.~A., et al.\ 2005, \apj, 628, 439 

\bibitem[Mitchell et al.(2003)]{2003AAS...203.1703M} Mitchell, D.~S., Frink, S., Quirrenbach, A., Fischer, D.~A., Marcy, G.~W., \& Butler, R.~P.\ 2003, Bulletin of the American Astronomical Society, 35, 1234 

\bibitem[Monnier et al.(2004)]{2004SPIE.5491.1370M} Monnier, J.~D., et al.\ 2004, \procspie, 5491, 1370

\bibitem[Naef et al.(2003)]{2003A&A...410.1051N} Naef, D., et al.\ 2003, \aap, 410, 1051 

\bibitem[Noyes et al.(1997)]{1997ApJ...483L.111N} Noyes, R.~W., Jha, S., Korzennik, S.~G., Krockenberger, M., Nisenson, P., Brown, T.~M., Kennelly, E.~J., \& Horner, S.~D.\ 1997, \apjl, 483, L111

\bibitem[Perrier et al.(2003)]{2003A&A...410.1039P} Perrier, C., Sivan, J.-P., Naef, D., Beuzit, J.~L., Mayor, M., Queloz, D., \& Udry, S.\ 2003, \aap, 410, 1039 

\bibitem[Perryman et al.(1997)]{1997A&A...323L..49P} Perryman, M.~A.~C., et al.\ 1997, \aap, 323, L49

\bibitem[Sato et al.(2003)]{2003ApJ...597L.157S} Sato, B., et al.\ 2003, \apjl, 597, L157 

\bibitem[Skrutskie et al.(2006)]{2006AJ....131.1163S} Skrutskie, M.~F., et al.\ 2006, \aj, 131, 1163 

\bibitem[Strassmeier \& Rice(2003)]{2003A&A...399..315S} Strassmeier, K.~G., \& Rice, J.~B.\ 2003, \aap, 399, 315 

\bibitem[Stepinski \& Black(2001)]{2001A&A...371..250S} Stepinski, T.~F., \& Black, D.~C.\ 2001, \aap, 371, 250 

\bibitem[ten Brummelaar et al.(2005)]{2005ApJ...628..453T} ten Brummelaar, T.~A., et al.\ 2005, \apj, 628, 453

\bibitem[Tsukamoto \& Makino(2007)]{2007ApJ...669.1316T} Tsukamoto, Y., \& Makino, J.\ 2007, \apj, 669, 1316 

\bibitem[Vogt et al.(2000)]{2000ApJ...536..902V} Vogt, S.~S., Marcy, G.~W., Butler, R.~P., \& Apps, K.\ 2000, \apj, 536, 902

\bibitem[Wu et al.(2007)]{2007ApJ...670..820W} Wu, Y., Murray, N.~W., \& Ramsahai, J.~M.\ 2007, \apj, 670, 820

\end{thebibliography}
\end{document}